\documentclass[11pt,twoside]{article}
\usepackage{asp2014}

\aspSuppressVolSlug
\resetcounters

\begin{document}

\title{Astronomical data organization, management and access in Scientific Data Lakes}

\author{Y.~G.~Grange$^1$, V.~N.~Pandey$^1$, X.~Espinal$^2$, R.~Di~Maria$^2$, and A.~P.~Millar$^3$, on behalf of ESCAPE WP2}
\affil{$^1$ASTRON, the Netherlands Institute for Radio Astronomy, Oude Hoogeveensedijk 4, 7991 PD, Dwingeloo, The Netherlands; \email{grange@astron.nl}}
\affil{$^2$CERN European Organization for Nuclear Research, Esplanade des particules 1, 1211 Geneva, Switzerland}
\affil{$^3$DESY, Notkestrasse 85, 22607 Hamburg, Germany}

\paperauthor{Yan~Grange}{grange@astron.nl}{0000-0001-5125-9539}{ASTRON}{Software Delivery}{Dwingeloo}{}{7991PD}{The Netherlands}
\paperauthor{Vishambhar~Nath~Pandey}{pandey@astron.nl}{0000-0001-8061-3760}{ASTRON}{Smart Backends}{Dwingeloo}{}{7991PD}{The Netherlands}
\paperauthor{Xavier~Espinal}{xavier.espinal@cern.ch}{0000-0001-7542-6098}{CERN}{}{Geneva}{}{1211}{Switzerland}
\paperauthor{Riccardo~Di~Maria}{riccardo.di.maria@cern.ch}{0000-0002-0186-3639}{CERN}{}{Geneva}{}{1211}{Switzerland}
\paperauthor{Paul~Millar}{paul.millar@desy.de}{0000-0002-3957-1279}{DESY}{}{Hamburg}{}{22607}{Germany}

\begin{abstract}
The data volumes stored in telescope archives is constantly increasing due to the development and improvements in the instrumentation. Often the archives need to be stored over a distributed storage architecture, provided by independent compute centres. Such a distributed data archive requires  overarching data management orchestration. Such orchestration comprises of tools which handle data storage and cataloguing, and steering transfers integrating different storage systems and protocols, while being aware of data policies and locality. In addition, it needs a common Authorisation and Authentication Infrastructure (AAI) layer which is perceived as a single entity by end users and provides transparent data access.

The scientific domain of particle physics also uses complex and distributed data management systems. The experiments at the Large Hadron Collider\,(LHC) accelerator at CERN generate several hundred petabytes of data per year. This data is globally distributed to partner sites and users using national compute facilities. Several innovative tools were developed to successfully address the distributed computing challenges in the context of the Worldwide LHC Computing Grid (WLCG). 

The work being carried out in the ESCAPE project and in the Data Infrastructure for Open Science (DIOS) work package is to prototype a Scientific Data Lake using the tools developed in the context of the WLCG, harnessing different physics scientific disciplines addressing FAIR standards and Open Data. We present how the Scientific Data Lake prototype is applied to address astronomical data use cases. We introduce the software stack and also discuss some of the differences between the domains.
\end{abstract}

\section{Introduction}
The ESCAPE project brings together the communities of astronomy, astroparticle and particle physics, as well as the relevant research infrastructures from the European Strategy Forum on Research Infrastructures (ESFRI). The project works towards the implementation of a functional link between the concerned projects and the European Open Science Cloud (EOSC).

Work Package\,2\,(WP2) of ESCAPE, called Data Infrastructures for Open Science\,(DIOS), aims to assess how the infrastructure, that has been in use for managing the data for several LHC experiments, commonly known as the Scientific Data Lake, could be applicable to the broader scope of the facilities within ESCAPE. Apart from the use cases that produce large amounts of data which needs to be managed, this also includes projects producing more moderate amounts of data. Those are the projects that need to ensure that data is replicated at the appropriate level, and that perform processing in a distributed fashion, on different compute environments.


\section{Scientific Data Lake}

The Scientific Data Lake infrastructure consists of several services that work together to manage the large data volumes from the LHC experiments as briefly illustrated in Fig.\,\ref{fig:datalake}.  It handles heterogeneity of storage systems by design, supporting endpoints with different file systems managed by different external parties including supporting different access protocols. This also allows for integration with heterogeneous computing resources, like grid-like sites, high-performance computers, commercial, private and hybrid clouds as well as sporadic resources.  The Scientific Data Lake helps in making experimental data {\it FAIR} \citep{Wilkinson2016}\footnote{\url{https://www.go-fair.org/fair-principles/}}. {\em Findable} and {\em Accessible} through a common name space,  {\em Interoperable} and {\em Reusable} by supporting easy transfers to various processing systems.  

\articlefigure{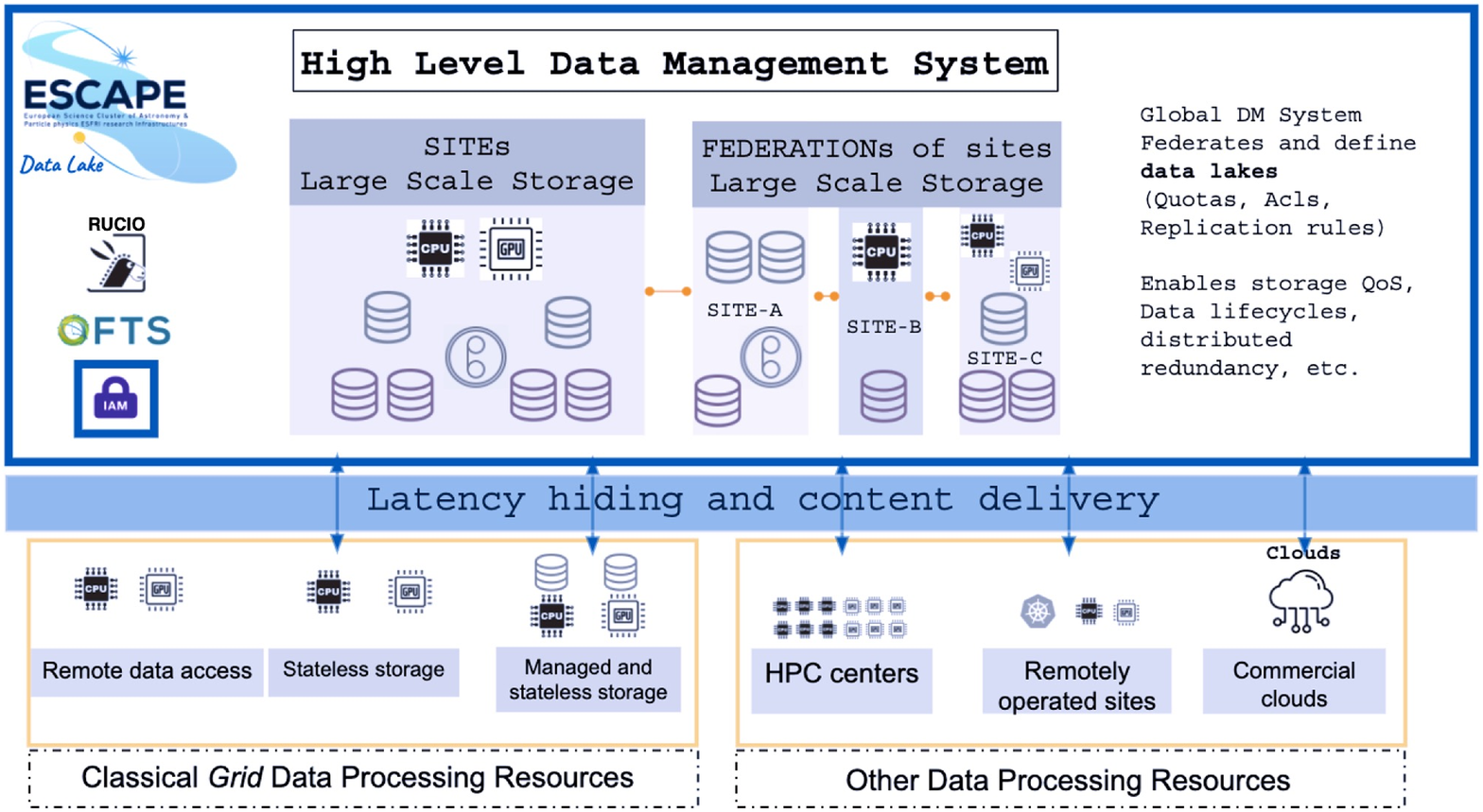}{fig:datalake}{A brief schematic overview of the Scientific Data Lake. In the top part, the three main pillars Rucio, FTS and IAM are shown, next to an illustration of a distributed set of storage end points, each with their own storage file systems (e.g. eos, dCache) and hardware (e.g. disk, tape). Some sites may also offer processing facilities. The bottom part shows different types of processing environment. The left half represents the classical Grid infrastructure, used to process the data from the LHC experiments. The right half lists other broadly adopted processing systems, like High-Performance Computing (HPC) or clouds. The content delivery and latency hiding layer of the infrastructure ensures that data can be accessed from any of those.}

\noindent{The architecture is built on three main pillars consisting of\,:}
\begin{description}
    \item [\textbf{Rucio}] Rule-based data management \citep{Barisits2019}. Rucio brings files and data sets that are stored in multiple locations under the same name space, and makes it possible to apply fine-grained data management concepts like polices, replication rules, data life-cycles, etc.
    \item[\textbf{FTS}] File Transfer Service \citep{Ayllon2014} is currently responsible for globally distributing the majority of Large Hadron Collider (LHC) data. It takes care of reliably transferring data between two endpoints, taking care of lower-level issues like matching protocols or retries. 
    \item[\textbf{Indigo IAM}] Identity and Access Management \citep{Ceccanti2018} provides the AAI capabilities. Users can be authenticated via their institutional credentials (e.g. through EduGAIN) and user groups can be managed which are provided to the underlying applications (like Rucio and FTS) so that access rights can be appropriately applied to the group members.
\end{description}

\section{Science Data Lake and astronomy \label{sec:dla}}
One of the main goals of ESCAPE WP2 is to assess the suitability of the Scientific Data Lake infrastructure  to support work flows from the astronomical use cases, suggest possible improvements and required new functionalities to the developers of the tooling. We give a brief overview of some of the main differences  between the data work flows in both fields including  ongoing relevant work towards addressing them. 

Different metadata for files. Rucio currently offers full support for file metadata (size, checksum, etc.) and metadata keys that are accelerator physics-specific. Custom metadata can be provided through a plugin or JSON file. It is currently not possible to automatically apply rules based on this custom metadata. Also the rule engine presently only supports (partial) equality comparison of metadata keys. Work is currently being done to extend the metadata-based rule execution to custom metadata, and extending the rule engine to support inequalities as well in queries (e.g. 01h42m30s<RA<01h45m30s).

In particle physics, access to data is generally granted based on X509 certificates. Obtaining a certificate for access can however be cumbersome, as it requires a trusted party to confirm ones identity when requesting, and the certificate needs to be regularly refreshed (typically every year). During the project, support for OpenID Connect (OIDC)\footnote{\url{https://openid.net/}} tokens for AAI is being implemented and activated. This makes it possible for user authenticcation  using home institute credentials resulting in  a more accessible system for non-expert users as well. Currently, the ESCAPE WP2 Rucio instance now supports usage of OIDC tokens. However, some of the the storage system back-ends still require X509 certificates for data access (i.e. downloads).

Between astronomy and particle physics, the data access models significantly differ. While in particle physics user access is centrally managed by a project, in astronomy not only data access must be provided by the observatory, but additionally it should also be possible for the PI of the project to be able to decide on granting access to the data. In addition the concepts of proprietary and public data, which are very common in astronomy but not so in particle physics has implications on AAI.

Non-expert users represent a large fraction of public data users in astronomy, while in particle physics the data is generally accessed by experts. To improve the data access by non-expert users, a Jupyter Lab kernel that provides easy access to the Scientific Data Lake (called DataLake as a Service; DLaaS) has been developed within the project.

\section{Conclusion and Future work}
In the first phase of the ESCAPE project, we have successfully assessed how the Scientific Data Lake architecture can be used to manage data from a wide range of experiments from the fields of fields of (astro)particle physics and astronomy. 

In the next phase, further integration between the Scientific Data Lake and the other components developed in ESCAPE will be one of the focal points. Further work is ongoing towards integration the DLaaS and the Science Analysis Platform developed in WP5 of ESCAPE \citep[see][for an overview of this platform]{X0-010_adassxxxi}. Querying the Scientific Data Lake as one of the data sources in the Science Analysis Platform makes it possible to combine data sets  from different collections or to customise query functionality for a specific field, for instance by offering astronomy-specific query parameters to astronomers. Also, there is an opportunity to investigate how the Scientific Data Lake could be integrated with the VOSpace standard \citep{Graham2018}, easing the integration of the Scientific Data Lake in the Virtual Observatory. Also the the support for both public and proprietary data  and the granular group management, as both being mentioned in Sect. \ref{sec:dla}, needs further investigation. Initial work to demonstrate ability to group scientifically related astronomical observations (e.g. calibrators and target observations etc.) using the concepts of  datasets, containers,and hierarchy on the scientific data lake and utilize them in processing work flows has shown promising results and thus needs further investigation.

Several project partners (MAGIC+CTA, SKAO and LSST) have set up their own instance for further exploration and ASTRON is planning to do so as well. This makes it possible to assess the architecture from the operational level, and also reap the gains made past the time frame of the ESCAPE project. 

\acknowledgements ESCAPE European Science Cluster of Astronomy \& Particle physics ESFRI research Infrastructures has received funding from the European Union's Horizon 2020 research and innovation programme under the Grant Agreement n$^\mathrm{o}$ 824064.

\bibliography{X9-002}


\end{document}